\documentclass[conference]{IEEEtran}
\pdfoutput=1
\ifCLASSINFOpdf
\else
\fi
\usepackage{epsfig, graphicx}
\usepackage{url}
\usepackage{hyperref}
\begin{document}
%
\title{Green Cellular - Optimizing the Cellular Network for Minimal Emission from Mobile Stations}

\author{\IEEEauthorblockN{Doron Ezri}
\IEEEauthorblockA{Greenair Wireless\\
Ramat Gan, Israel\\
Email: doron@greenairwireless.com}
\and
\IEEEauthorblockN{Shimi Shilo}
\IEEEauthorblockA{Greenair Wireless\\
Ramat Gan, Israel\\
Email: shimi@greenairwireless.com}}


%


\maketitle

\begin{abstract}
Wireless systems, which include cellular phones, have become an essential part of the modern life.  However the  mounting evidence that cellular radiation might adversely affect the health of its users, leads to a growing concern among authorities and the general public. Radiating antennas in the proximity of the user, such as antennas of mobile phones are of special interest for this matter. In this paper we suggest a new architecture for wireless networks, aiming at minimal emission from mobile stations, without any additional radiation sources. The new architecture, dubbed Green Cellular, abandons the classical transceiver base station design and suggests the augmentation of transceiver base stations with receive only devices. These devices, dubbed Green Antennas, are not aiming at coverage extension but rather at minimizing the emission from mobile stations. We discuss the implications of the Green Cellular architecture on 3G and 4G cellular technologies. We conclude by showing that employing the Green Cellular approach may lead to a significant decrease in the emission from mobile stations, especially in indoor scenarios. This is achieved without exposing the user to any additional radiation source.
\end{abstract}


%
\IEEEpeerreviewmaketitle

\section{Introduction}
With over 3 billion cellular users worldwide, cellular technology is literally everywhere. Thus, the significant effort to figure out the implications of cellular usage on human health is no surprise, nor is the interest of authorities and the public in this issue.

The precautionary advice on cell phone usage as published by the University of Pittsburgh's Cancer Institute \cite{Pittsburgh} is one of the most influencing statements regarding cell phone safety in recent years. This statement lists ten precautions aiming at reducing exposure to cellular radiation. Special attention is given to young children who are considered most likely to be sensitive to exposure. A similar concern has led French authorities to ban cell phone usage in primary schools \cite{FrenchPrimary}. On the research front, perhaps the most thorough study set up to investigate whether cell phone use increases the risk of cancer, in the multinational INTERPHONE Study \cite{Interphone}. So far no consensus has been reached, but some of the intermediate results link cell phone usage to certain types of cancer \cite{Sadetzki}.

It seems that as cellular and wireless technologies evolve, the exposure to radiation only grows. Application of modern cellular and wireless technologies are ubiquitous, and are by no means limited to voice services. Popular smart phones lead to an increased  usage profile (e.g. web browsing, social networks, emailing) \cite{comScore} and  bandwidth hungry applications such as video streaming are becoming more prominent, leading to increased exposure.

Current solutions aiming at reducing cellular radiation from mobile stations (MSs) are roughly divided into the following types. The first is an increase in the number of cellular sites, leading to a reduction in the average separation between the site and the MS (smaller cells). This approach leads to a reduction in the average emission of both MSs and cellular sites. An immediate drawback here is the concern of nearby residents from constant radiation from the cellular site. Moreover, this approach gives a partial solution to indoor users who may still experience high levels of emission due to significant penetration loss \cite{PenetrationLossHoppe,PenetrationLossMartijn}.

Another prominent solution is wired and wireless headsets. Focusing on wireless headsets (usually Class 2 Bluetooth devices), even though the average transmit (Tx) power of the headset (4dBm in Bluetooth), is lower than the average Tx power of an MS, in many occasions (especially with the high dynamic range of 3G and 4G cellular) the MS itself may transmit at lower power than the bluetooth transmitter. Furthermore, the MS is always kept nearby, usually worn on the body of the user, implying additional exposure to radiation. The impact of wired headsets is also debatable as reflected from contradicting findings \cite{Troulis,BitBabik}.

Finally, femtocells and indoor distributed antenna systems (DASs), which are a limit case of the smaller cell solution, may be viewed as a means to reduce radiation. Obviously, since these solutions are installed indoor, they result in a significant reduction in the MS' emission power. However, since they transmit at a non negligible power \cite{Chandrasekhar2008}, they give rise to concerns similar to those respective to smaller cells.

In this paper we do not address the possible health implications of cellular radiation,  but rather adopt the precautionary principle which guides us to minimize exposure to radiation \cite{Precautionary,LinRose}. Current cellular architectures are adapted to meet other design criteria, such as coverage and capacity. In this paper we address this problem and propose a new architecture for cellular and wireless networks which is optimized for minimal emission from MSs, without any additional radiation sources. We further give experimental results showing the significant effect the proposed architecture has on MS emission power.

The paper is organized as follows. In Section 2 we present the new architecture for cellular and wireless networks dubbed Green Cellular. In Section 3 we discuss the implications on Green Cellular of 3G and 4G technologies.  In Section 4 we give simulation results showing the significant effect the Green Cellular architecture has on MS emission power. Section 5 contains discussion and conclusions.

\section{The Green Cellular Architecture}

Most modern cellular and wireless networks (e.g. 3G and 4G technologies), employ sophisticated power control mechanisms, controlling the Tx power of both base stations (BSs) and MSs. These power control mechanism underly the small cell approach that lead to lower Tx power at both BSs and MSs. Focusing on the uplink (UL), we emphasize that the MS Tx power depends specifically on the separation from BS \emph{receivers} (and not necessarily transceivers).

\begin{figure}[h]
\centering
\resizebox{!}{4cm}{\includegraphics{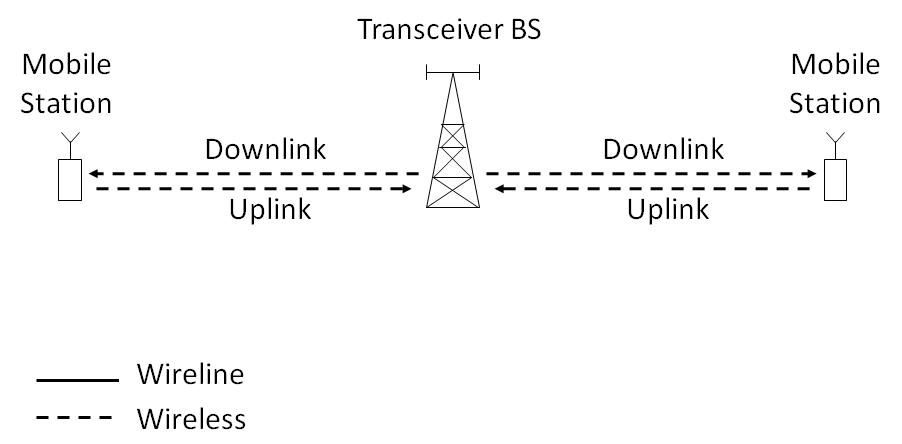}}
\caption{\label{RegularScheme}Standard transceiver architecture.}
\end{figure}

\begin{figure}[h]
\centering \resizebox{!}{4cm}{\includegraphics{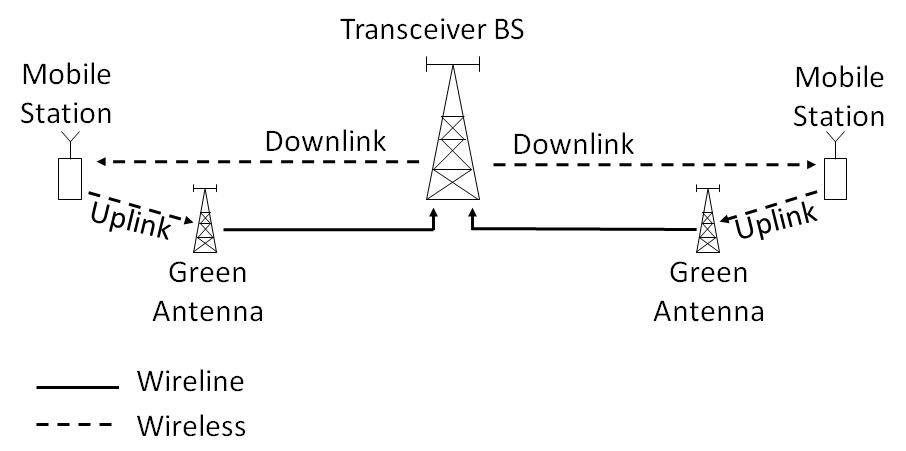}}
\caption{\label{GreenScheme}The Green Cellular architecture in which transceiver BSs are augmented with Green Antennas. The UL to the BS is not shown.}
\end{figure}

While most wireless and cellular networks adopt an architecture that is based on transceiver BSs, the Green Cellular approach suggests the augmentation of transceiver BSs with receive only devices. The latter, dubbed Green Antennas, are employed to reduce the Tx power of MSs in their proximity\footnote{The concept of Green Cellular as well as respective architecture and technological implications have been disclosed in a series of patent applications \cite{OurPatent1,OurPatent2,OurPatent3,OurPatent4}.}. In order to minimize exposure to cellular radiation, the Green Antennas are to be connected to the network infrastructure via wireline or highly directional point-to-point  microwave link. Theoretically, one may deploy Green Antennas on a sufficiently dense grid, reducing the average Tx power of MSs to any arbitrary value (supported by the MS), without any additional radiation source. Green Cellular is a broad concept that may be applied to a wide array of technologies ranging from small local area networks (LANs) to large  metropolitan area networks (MANs). A schematic of the Green Cellular architecture is given in Fig. \ref{GreenScheme}. Compare with the standard architecture based on transceiver BS, given in Fig. \ref{RegularScheme}.


Focusing on cellular MANs, bearing in mind significant cellular usage is indoors \cite{MobileCallsIndoors}, a natural embodiment of Green Cellular is the augmentation of outdoor BSs with indoor Green Antennas. Due to the relatively high MS Tx power indoors (mainly stemming from penetration loss), Green antennas installed indoors play an important role in this architecture. For example, indoor Green Antennas may be installed within buildings where radiation is a major concern, such as schools. This way, the outdoor BS may be located at any distance from the school, still allowing admissible coverage.

MSs in the proximity of a Green Antenna would transmit at significantly lower Tx power (compared to the same scenario lacking the Green Antenna), reducing the interference to users of both the same BS and neighboring BSs. Since most practical communication systems are interference limited (see \cite{Goldsmith}, Page 507), this would also result in a decrease of the required Tx power of other MSs (not at an immediate proximity of a Green Antenna).

The Green Cellular architecture provides  further enhancements beyond the decrease in MS Tx power. Green Antennas lead to significantly higher UL capacity both at the serving and neighboring BSs (similar to femtocells \cite{Chandrasekhar2008}),  an issue that is becoming more important as usage profile creates more demand for UL traffic. The significantly lower Tx power results in prolonged MS battery life. Finally, easier to deploy since they are expected to impose no regulatory constraints, dense deployment of Green Antennas would  enhance location based services (see \cite{LBS_Book}, Chapters 8 and 11).

\section{Technological Implications}

Practical communication systems have limitations such as limited dynamic range, degenerate power control mechanism etc. Note that the Green Cellular architecture implicitly assumes closed-loop UL power control (see \cite{WCDMA_UMTS}, Page 48), since the UL and downlink (DL) channels may significantly differ. This does not impose a serious constraint as most modern cellular and wireless technologies accommodate closed-loop power control with a substantial dynamic range. For example, the dynamic range in UMTS is over 60dB (see \cite{WCDMA_UMTS}, Page 326).

In systems where diversity combining, taking signal-to-interference-plus-noise-ratio (SINR) into consideration, is applied, the employment of Green Antennas is simple as they can only enhance the performance of the system. Such combining methods are maximal ratio combining (MRC), log likelihood ratios (LLRs) summation and combining of cyclic redundancy check (CRC) confirmed packets  (see \cite{Sesia}, Page 563). These methods are prominent in orthogonal frequency division multiplexing (OFDM) and OFDM-like technologies (e.g. WiMAX and LTE).

In case simple combining methods are employed (e.g. equal gain combining or analog combining), Green Antennas should be placed taking network topology into consideration. For instance, a Green Antenna associated with a certain BS, should not be placed within the dominant coverage area of neighboring BS (e.g. closer to the neighboring BS). This may lead to the Green Antenna receiving mostly signals from users of the neighboring BS, which results in an increased interference.

\begin{figure}[h]
\centering \resizebox{!}{4cm}{\includegraphics{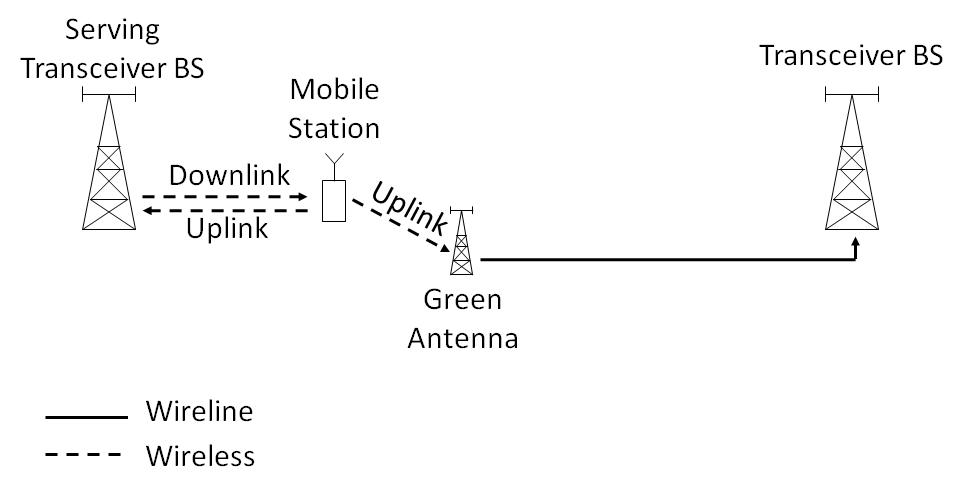}}
\caption{\label{BadPositioning}A scenario where the MS does not connect to the BS with nearest Green Antenna.}
\end{figure}

In most practical systems, the MS requests to connect to the BS which it receives with highest power or SINR (see \cite{WCDMA_UMTS}, Page 167). When such systems are augmented with Green Antennas, the MS will connect to the BS which it receives the strongest, which is not necessarily the BS which receives the MS with best quality (e.g. the BS with nearest Green Antenna). This scenario is given in Fig. \ref{BadPositioning}. This problem may be overcome by connecting the Green Antenna to all dominant neighboring BSs, altering the access and handover mechanisms to accommodate Green Antennas, or proper cell planning.

\section{Simulation Results}

In order to investigate the impact of Green Antennas on the Tx power of nearby MSs, a simulation study was conducted. The simulation is based on a commercial\footnote{We used a 4G simulation tool by SCHEMA Ltd.} 4G cellular network planning and optimization tool that has been adapted to accommodate Green Antennas. The Green Antennas forward the received signal from each MS to its respective serving transceiver BS. In this simulation we consider outdoor Green Antennas only.

The network simulation and planning tool used is optimized for real network configurations of urban areas.
A mixed scenario was used for the simulation, in which voice and data traffic distributions as well as indoor/outdoor user distributions were taken into account.
Penetration loss, lognormal fading, user speed distribution and other parameters were defined per clutter types according to typical values for an urban area. Relatively high traffic load was assumed such that the UL is interference limited and closed-loop UL power control was activated.
A Green Antenna was added into an area of relatively poor coverage. The impact of the Green Antenna on the MSs' uplink Tx power was analyzed in an area approximately 300m in diameter around the location of the Green Antenna. Two scenarios have been examined. In the first a regular deployment of transceiver BSs was assumed. In the second, Green Antennas were introduced.

\begin{figure}[h]
\centering \resizebox{!}{7cm}{\includegraphics{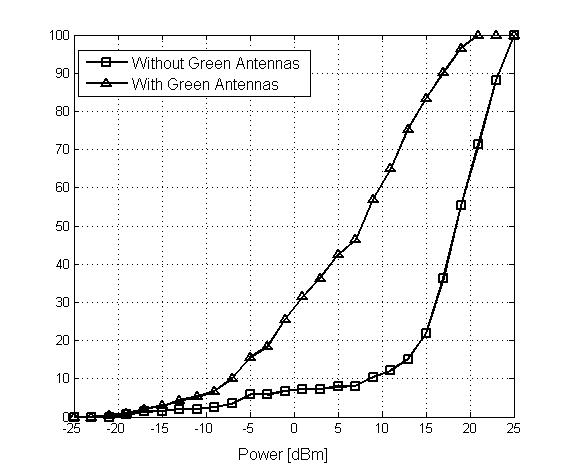}}
\caption{\label{CDF_TxPower}Comparison of the CDF of the MSs Tx power with and without Green Antennas. }
\end{figure}

 The complementary distribution function (CDF) of indoor MSs' Tx power in a radius of 300m about the Green Antennas in both scenarios is given in Fig. \ref{CDF_TxPower}. The figure shows a significant 8dB difference in the average Tx power and 10dB difference in the median. Considering a target Tx power of 4dBm (the Tx power of a Class 2 Bluetooth headset), note that in the scenario without Green Antennas only 7\% of the users are below the target power, whereas approx. 40\% are below the target with Green Antennas.

\section{Discussion and Conclusions}

In this paper we proposed a new architecture for cellular and wireless systems. Perhaps the most interesting aspect of the Green Cellular architecture is its ability to significantly reduce the emission from MSs without any additional radiation sources. Our simulation results showed over 10dB decrease in average Tx power of indoor MSs in the proximity of an outdoor Green Antenna. The effect of indoor installed Green Antennas is expected to be significantly more dramatic due to lower penetration losses (the effect is expected to be orders of magnitude more substantial with indoor Green Antennas). The new architecture does not require a reorganization of the network but rather implies an augmentation of the existing network with Green Antennas. Viewing Green Antennas as remotely installed diversity antennas makes this augmentation natural in systems employing SINR aware diversity combining.

The precautionary principle guides us to design cellular and wireless systems that minimize exposure to radiation. The Green Cellular concept described in this paper is perfectly aligned with this principle. The concept that the Tx power from MSs can be arbitrarily low, brought about by augmenting the network with Green Antennas, opens up a new realm of embodiments which are expected to play an important role in future communication systems.

More advanced considerations arising from the Green Cellular architecture, such as the decrease in overall UL interference, indoor installed Green Antennas' performance gains, increase in cell capacity and access/handover methods accommodating Green Antennas will be addressed in forthcoming papers.

\section*{Acknowledgement}
The authors wish to thank Prof. Shlomo Shamai (Shitz) for helpful discussions and Dr. Michael Livschitz of SCHEMA for conducting the simulations.

\bibliographystyle{IEEEtran}
\bibliography{Greenair_bib220609}

\begin{thebibliography}{10}
\providecommand{\url}[1]{#1}
\csname url@samestyle\endcsname
\providecommand{\newblock}{\relax}
\providecommand{\bibinfo}[2]{#2}
\providecommand{\BIBentrySTDinterwordspacing}{\spaceskip=0pt\relax}
\providecommand{\BIBentryALTinterwordstretchfactor}{4}
\providecommand{\BIBentryALTinterwordspacing}{\spaceskip=\fontdimen2\font plus
\BIBentryALTinterwordstretchfactor\fontdimen3\font minus
  \fontdimen4\font\relax}
\providecommand{\BIBforeignlanguage}[2]{{%
\expandafter\ifx\csname l@#1\endcsname\relax
\typeout{** WARNING: IEEEtran.bst: No hyphenation pattern has been}%
\typeout{** loaded for the language `#1'. Using the pattern for}%
\typeout{** the default language instead.}%
\else
\language=\csname l@#1\endcsname
\fi
#2}}
\providecommand{\BIBdecl}{\relax}
\BIBdecl

\bibitem{Pittsburgh}
R.~{Herberman}, ``{Tumors and cell phone use: what the science says},''
  Website, Sept. 2008,
  \url{http://domesticpolicy.oversight.house.gov/documents/20080925142803.pdf}.

\bibitem{FrenchPrimary}
C.~{Bremner}, ``{Mobiles phones to be banned in French primary schools to limit
  health risks},'' Website, May 2009,
  \url{http://www.timesonline.co.uk/tol/news/world/europe/article6366590.ece}.

\bibitem{Interphone}
{E. Cardis, \emph{et al.}}, ``{The INTERPHONE study: design, epidemiological
  methods, and description of the study population},'' \emph{{Eur. J.
  Epidemiol.}}, vol.~22, pp. 647--664, July 2007.

\bibitem{Sadetzki}
\BIBentryALTinterwordspacing
{S. Sadetzki, A. Chetrit, A. Jarus-Hakak, E. Cardis, Y. Deutch, S. Duvdevani,
  A. Zultan, I. Novikov, L. Freedman, and M. Wolf}, ``Cellular phone use and
  risk of benign and malignant parotid gland tumors -- a nationwide case -
  control study,'' \emph{Am. J. Epidemiol.}, vol. 167, no.~4, pp. 457--467,
  Feb. 2008. [Online]. Available: \url{http://dx.doi.org/10.1093/aje/kwm325}
\BIBentrySTDinterwordspacing

\bibitem{comScore}
{J. Steele}, ``{comScore: mobile internet becoming a daily activity for
  many},'' Website, March 2009,
  \url{http://www.comscore.com/Press_Events/Press_Releases/2009/3/Daily_Mobile%
_Internet_Usage_Grows}.

\bibitem{PenetrationLossHoppe}
{R. Hoppe, G. Wolffe, and F. M. Landstorfer}, ``{Measurement of building
  penetration loss and propagation models for radio transmission into
  buildings},'' \emph{IEEE Veh. Technol. Conf.}, pp. 2298--–2302, Sept. 1999.

\bibitem{PenetrationLossMartijn}
{E. F. T. Martijn, and M. H. A. J. Herben}, ``{Characterization of radio wave
  propagation into buildings at 1800 MHz},'' \emph{IEEE Ant. and Wireless Prop.
  letters}, vol.~2, pp. 122--125, 2003.

\bibitem{Troulis}
{S. E. Troulis, W. G. Scanlon, and N. E. Evans}, ``{Effect of a hands-free wire
  on specific absorption rate for a waist-mounted 1.8 GHz cellular telephone
  handset},'' \emph{Physics in Medicine and Biology}, vol.~48, pp. 1675--1684,
  2003.

\bibitem{BitBabik}
{G. Bit-Babik, C. K. Chou, A. Faraone, A. Gessner, M. Kanda, and Q. Balzano},
  ``{Estimation of the SAR in the human head and body due to radiofrequency
  radiation exposure from handheld mobile phones with hands-free
  accessories},'' \emph{Radiation Research}, vol. 159, pp. 550--557, 2003.

\bibitem{Chandrasekhar2008}
{V. Chandrasekhar, J. G. Andrews, and A. Gatherer}, ``{Femtocell networks: A
  survey},'' \emph{IEEE Comm. Magazine}, vol.~46, no.~9, pp. 59--67, Sept.
  2008.

\bibitem{Precautionary}
{I. K. Leeka, L. H. Gordon, and L. B. Gail}, ``{The precautionary principle and
  EMF: implementation and evaluation},'' \emph{Journal of Risk Research},
  vol.~4, no.~2, pp. 113--125, April 2001.

\bibitem{LinRose}
{J. C. Lin}, ``{The precautionary principle - rose by another name},''
  \emph{IEEE Antennas and Prop. Magazine}, vol.~43, no.~2, pp. 129--131, Sept.
  2001.

\bibitem{OurPatent1}
{D. Ezri}, ``{Wireless communication system, device and method},'' \emph{U.S
  Patent Application no. 61/119,755}, 2009.

\bibitem{OurPatent2}
------, ``{Wireless communication system, device and method},'' \emph{U.S
  Patent Application no. 61/120,458}, 2009.

\bibitem{OurPatent3}
------, ``{System and method for connection and handover in wireless
  communication},'' \emph{U.S Patent Application no. 61/121,595}, 2009.

\bibitem{OurPatent4}
{D. Ezri, and S. Shilo}, ``{Wireless communication system, device and
  method},'' \emph{U.S Patent Application no. 61/175,063}, 2009.

\bibitem{MobileCallsIndoors}
{C. Wei, and B. E. Kolko}, ``{Studying mobile phone use in context: cultural,
  political, and economic dimensions of mobile phone use},'' \emph{Int. Prof.
  Comm. Conf.}, pp. 205 -- 212, 2005.

\bibitem{Goldsmith}
{A. Goldsmith}, \emph{{Wireless Communications}}.\hskip 1em plus 0.5em minus
  0.4em\relax Cambridge University Press, 2005.

\bibitem{LBS_Book}
{A. K\"upper}, \emph{{Location-Based Services: Fundamentals and
  Operation}}.\hskip 1em plus 0.5em minus 0.4em\relax {Wiley}, 2005.

\bibitem{WCDMA_UMTS}
{H. Holma, and A. Toskala}, \emph{{WCDMA for UMTS}}.\hskip 1em plus 0.5em minus
  0.4em\relax {John Wiley}, 2002.

\bibitem{Sesia}
{S. Sesia, I. Toufik, and M. Baker}, \emph{{LTE, The UMTS Long Term Evolution:
  From Theory to Practice}}.\hskip 1em plus 0.5em minus 0.4em\relax {Wiley},
  2009.

\end{thebibliography}

\end{document}